\documentclass[10pt]{iopart}
\usepackage{iopams}  
\usepackage{graphicx}

\begin{document}
\title[Dip-coating with prestructured substrates: transfer of simple liquids and Langmuir-Blodgett monolayers]{Dip-coating with prestructured substrates: transfer of simple liquids and Langmuir-Blodgett monolayers}

\author{Markus Wilczek$^1$, Juan Zhu$^2$, Lifeng Chi$^{3}$, Uwe Thiele$^{1,4}$, Svetlana V. Gurevich$^{1,4}$}

\address{1 Institute for Theoretical Physics, University of M\"unster, Wilhelm-Klemm-Str.\ 9, D-48149 M\"unster, Germany; Center for Nonlinear Science (CeNoS), University of M\"unster, Corrensstr.\ 2, D-48149 M\"unster,  Germany}
\address{2 Physical Institute, University of M\"unster, Wilhelm-Klemm-Str.\ 10, D-48149 M\"unster, Germany; 
Center for Nanotechnology (CeNTech), University of M\"unster, Heisenbergstr.\ 11, D-48149 M\"unster, Germany}
\address{3 Institute of Functional Nano \& Soft Materials (FUNSOM) and Jiangsu Key Laboratory for Carbon-based Functional Materials \& Devices; Collaborative Innovation Center of Suzhou Nano Science and Technology, Soochow University, 199 Ren-ai Road, Suzhou 215123, P. R. China}
\address{4 Center for Multiscale Theory and Computation (CMTC), University of M\"unster, Corrensstr.\ 40, D-48149 M\"unster, Germany}
\ead{markuswilczek@uni-muenster.de}

\begin{abstract}
When a plate is withdrawn from a liquid bath, either a static meniscus forms in the transition region between the bath and the substrate or a liquid film of finite thickness (a Landau-Levich film) is transferred onto the moving substrate. If the substrate is inhomogeneous, e.g., has a prestructure consisting of stripes of different wettabilities, the meniscus can be deformed or show a complex dynamic behavior. Here we study the free surface shape and dynamics of a dragged meniscus occurring for striped prestructures with two orientations, parallel and perpendicular to the transfer direction.  A thin film model is employed that accounts for capillarity through a Laplace pressure and for the spatially varying wettability through a Derjaguin (or disjoining) pressure. Numerical continuation is used to obtain steady free surface profiles and corresponding bifurcation diagrams in the case of substrates with different homogeneous wettabilities.  Direct numerical simulations are employed in the case of the various striped prestructures. \par
The final part illustrates the importance of our findings for particular applications that involve complex liquids by modeling a Langmuir-Blodgett transfer experiment. There, one transfers a monolayer of an insoluble surfactant that covers the surface of the bath onto the moving substrate. The resulting pattern formation phenomena can be crucially influenced by the hydrodynamics of the liquid meniscus that itself depends on the prestructure on the substrate. In particular, we show how prestructure stripes parallel to the transfer direction lead to the formation of bent stripes in the surfactant coverage after transfer and present similar experimental results.
\end{abstract}

%
\vspace{2pc}
\maketitle
%
\ioptwocol

\section{Introduction}
Many experimental set-ups as well as industrial applications involve the transfer of simple and complex liquids onto solid substrates by withdrawing the substrates from a liquid bath at a certain velocity \cite{WeRu2004arfm,SADF2007jfm,DFSA2008jfm,Wils1982jem}. Examples are dip-coating processes with simple liquids, solutions or suspensions (e.g., \cite{li2010controllable,li2013growth,doumenc2013numerical,corrales2014spontaneous,wang2015addressable}) or the Langmuir-Blodgett transfer process that is used to transfer monomolecular layers of amphiphilic molecules from the surface of a liquid bath onto a solid (e.g., \cite{GlCF2000n}). In such set-ups, the liquid-gas interface forms a meniscus close to the three-phase contact line. The meniscus may be static when capillary forces stemming from surface tension, wettability, i.e., solid-liquid interactions, as well as the viscous drag due to the moving substrate balance. Among the best known phenomena, however, is not the formation of a meniscus but the transfer of a liquid layer of finite thickness onto the substrate that occurs above a critical transfer velocity (see, e.g., \cite{LaLe1942apu,MRQG2007l}). This was first analyzed by Landau and Levich \cite{LaLe1942apu}, and therefore such a liquid film is widely known as Landau-Levich film. Its thickness scales with the transfer velocity $U$ as $U^{\frac{2}{3}}$.\par
Here, we focus on the morphology of a liquid meniscus for transfer velocities below the transition to a Landau-Levich film. Interesting effects can be found already in this regime, foremost a transition from classical meniscus shapes with a monotonous change of the free surface slope to profiles that show \textsl{foot-like} structures, i.e., a liquid layer of finite height and length protrudes from the meniscus. This has been studied in detail using different models and approaches \cite{SADF2007jfm,ZiSE2009epjt,galvagno2014continuous,tseluiko2014collapsed,gao2016film}, also revealing the bifurcation structure of this system. Additional effects, such as the evaporation of the liquid and deposition of a solute, were also already discussed in the context of a dragged meniscus \cite{Dey2016}.\par
All the mentioned works only consider homogeneous substrates, i.e., substrates with completely homogeneous physical and chemical properties. However, in experiments this is not always the case, either naturally when considering realistic substrates that have impurities or defects, or when substrates are artificially structured for different reasons. We focus on such prestructured substrates that are, e.g., chemically treated in such a way that properties, such as the wettability, vary in a regular way across the substrate. Mostly, prestructures are used to achieve an additional control mechanism for the experimental processes. Examples are prestructured substrates used in Langmuir-Blodgett transfer or dip-coating experiments that are used to control the pattern formation processes occurring in these experiments \cite{DTDM2000jap,KoDi2004pf,WiGu2014pre,Zhu2016AMI,wang2015addressable}. Here, we will analyze the influence of such prestructures on the dynamics of a dragged liquid meniscus.

\begin{figure}[htbp]
\center
\includegraphics[width=0.45\textwidth]{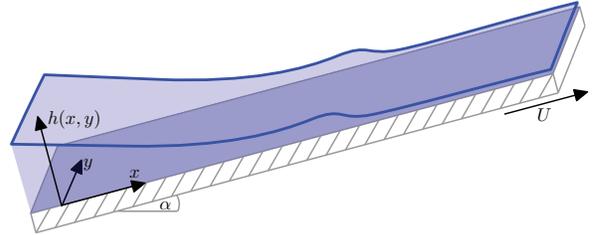}
\caption{Sketch of a meniscus that forms when a solid substrate  is withdraw from a liquid bath under an angle $\alpha$ with the velocity $U$. At the bottom left, our frame of reference is shown.}
\label{fig:sketch_dragged_meniscus}
\end{figure}

In particular, we are interested in the geometry sketched in Fig.~\ref{fig:sketch_dragged_meniscus}, where a substrate is dragged out from a liquid reservoir at a constant velocity $U$ and under a constant angle $\alpha$ with respect to the free surface  of the liquid reservoir. We use a thin-film or long-wave model in the form of an evolution equation for the film thickness profile $h(x,y,t)$ \cite{OrDB1997rmp,Thie2007} accounting for wettability, surface tension, hydrostatic pressure and the driving due to the transfer process. This model is analyzed using a numerical continuation technique \cite{Kuznetsov2010,DWCD2014ccp} to study two-dimensional (2d) physical situations, i.e., one-dimensional (1d) steady profiles $h(x)$ and obtain the bifurcation diagrams for the main control parameters: transfer velocity, inclination angle and substrate wettability. In addition, we use direct numerical simulations in two and three physical dimensions, i.e., for 1d [$h(x,t)$] and 2d [$h(x,y,t)$] profiles, respectively. This augments the bifurcation diagrams of steady 1d profiles with results for prestructured substrates that, depending on the particular prestructure, can be again steady profiles or show intrinsically dynamic behavior. A comparison of the results obtained with the different techniques shall enable us to base predictions for different types of prestructures on the bifurcation diagrams obtained for homogeneous substrates, without having to explicitly investigate all possible prestructure shapes.\par
Finally, we consider the Langmuir-Blodgett transfer of surfactant monolayers onto prestructured substrates to emphasize the relevance of our findings for real-world applications that involve complex liquids. To this end, we present both, results from experiments and simulation results obtained with a two-component thin-film model. We show that a good qualitative agreement is found and propose an explanation based on the findings presented before for simple liquids.\par  
This paper is organized as follows: In Sec.~\ref{sec:modelling} we introduce the thin-film model and the numerical approaches we use to analyze it. In Sec.~\ref{sec:bifurcation_diagram_v_length} we first recapitulate the findings from the literature in the case of homogeneous substrates and re-calculate them for our specific settings. Then we analyze the bifurcation diagram for varying substrate wettability in Sec.~\ref{sec:bifurcation_diagram_rho_length}. This forms the basis for the following Sec.~\ref{sec:heterogeneous}, where we consider spatially inhomogeneous substrates for two differently oriented stripe-like prestructures. In each case, direct numerical simulations are carried out and the results are connected to the bifurcation diagrams obtained for 1d homogeneous substrates. In Sec.~\ref{sec:LB} we expand our focus to the experimental set-up of Langmuir-Blodgett transfer onto prestructured substrates and present both, experimental and theoretical results. Finally, we conclude in Sec.~\ref{sec:summary}.

\section{Mathematical Modelling and Numerical Approach}
\label{sec:modelling}
To model the dynamics of the liquid layer forming the dragged meniscus, we use a non-dimensional thin film model that is derived from the Navier-Stokes equations and appropriate boundary conditions under the assumption of small gradients \cite{OrDB1997rmp,Thie2007}. It accounts for wettability through a Derjaguin (or disjoining) pressure $\Pi(h)$, capillarity through a Laplace pressure $-\Delta h$, the hydrostatic pressure $G h$, the driving induced by the transfer velocity $\mathbf{U} = (U,0)$ and the gravitational force parallel to the substrate $G\alpha$. Note that the used Derjaguin pressure allows for a coexistence of a thick liquid layer with a thin wetting layer, sometimes called a precursor film. 
The model describes the evolution of the 2d height-profile $h(x,y,t)$ (see Fig.~\ref{fig:sketch_dragged_meniscus} for a sketch clarifying the coordinate system):
\begin{eqnarray}
 \partial_t &h(x,y,t)\ &=  \nabla \cdot \Big( \frac{h^3}{3} \nabla \big(-\Delta h - \kappa \Pi(h)\big)   \nonumber \\
&  & \quad \quad\quad + \boldsymbol{\chi}(h) \Big)  -\mathbf{U} \cdot \nabla h, \label{TFEmodel}\\
&\Pi(h) \quad &=  - \frac{1}{h^3} + \frac{1}{h^6}, \\
&\boldsymbol{\chi}(h) \quad &= \frac{h^3}{3} G \big(\nabla h + \boldsymbol{\alpha} \big).
\end{eqnarray}
Here, $\boldsymbol{\alpha} = (\alpha,0)$ is the inclination angle of the substrate and $G$ a dimensionless gravity parameter. We refer to Refs.~\cite{galvagno2014continuous,tseluiko2014collapsed} for a detailed account of the scaling used to obtain the dimensionless equation. There, the model was already studied in detail in the one-dimensional case of homogeneous substrates. Special emphasis was put (i) on the bifurcation diagrams with the substrate velocity as the main control parameter and (ii) on foot-like solution profiles where a liquid layer of finite height and finite length connects the liquid bath and the thin wetting layer. Similar results have also be obtained with a slip model \cite{SADF2007jfm,ZiSE2009epjt}.\par 
An important aspect of the model and the special geometry considered are the boundary conditions at the downstream end $x=0$ of the domain $\Omega = [0,L_x]\times [0,L_y]$ (l.h.s.\ in Fig.~\ref{fig:sketch_dragged_meniscus}) where the free surface approaches the horizontal bath surface. There, the height of the liquid layer above the substrate asymptotically exhibits a linear behavior. Asymptotic boundary conditions at arbitrary order for finite domains are derived in \cite{tseluiko2014collapsed} through a central manifold approach. Here, however, we choose the domain sufficiently large to only use the lowest order terms. Effectively, at the downstream boundary $x=0$ we fix the liquid height at some $h_0$ and set the slope to $-\alpha$:
\begin{eqnarray}
 h|_{x = 0} = h_0, \quad \nabla h|_{x = 0} = (-\alpha,0), \quad \nabla h|_{x = L_x } = 0, \nonumber\\
 \nabla \Delta h|_{x = L_x } = 0, \quad
 h|_{y = 0} = h|_{y = L_y}. \label{eq:bc}
\end{eqnarray}
For the upstream boundary at $x = L_x$ we impose a flat layer by requesting the first and third derivative of $h$ to vanish. In the 2d case we impose periodic boundary conditions in $y$-direction.\par
In contrast to Refs.~\cite{galvagno2014continuous,tseluiko2014collapsed}, we re-introduce a parameter $\kappa$ that controls the wettability of the substrate. Strictly speaking, it represents the ratio of a particular wetting energy to the typical wetting energy employed in the scaling. In particular, $\kappa$ is related to the long-wave equilibrium contact angle by
\begin{equation}
\theta_\mathrm{eq} = \sqrt{ \frac{3\kappa}{5}}.
\end{equation}
The main objective of the present work is to study the influence of varying wettability $\kappa$, first in the case of spatially homogeneous substrates. Second, the obtained results are related to the case of prestructured substrates where $\kappa(x,y)$ is spatially varying about the reference value, i.e., about $\kappa=1$. In the latter case we assume prestructures that correspond either to stripes parallel to the transfer velocity
\begin{equation}
\label{prestructure_y}
\kappa(y) = 1 + \rho \tanh\left[s \left(4 \left|\mathrm{frac}\left(\frac{y}{L_\mathrm{pre}}\right)-0.5\right|-1\right)\right]
\end{equation}
or perpendicular to the transfer velocity
\begin{equation}
\label{prestructure_x}
\kappa(x) = 1 + \rho \tanh\left[s \left(4 \left|\mathrm{frac}\left(\frac{x-Ut}{L_\mathrm{pre}}\right)-0.5\right|-1\right)\right].
\end{equation}
In both cases, $\mathrm{frac}$ denotes the fractional part of the argument and the prestructure has a periodicity $L_\mathrm{pre}$, a steepness $s$ and a contrast $\rho$, indicating that the wettability varies between $\kappa = 1- \rho$
and $\kappa = 1+\rho$. \par
To perform direct numerical simulations of the model (\ref{TFEmodel}), we employ a finite-element method, implemented using the open source framework DUNE-PDELab \cite{bastian2010generic,bastian2008genericI,bastian2008genericII}. The 2d simulation domain $\Omega = [0,L_x]\times [0,L_y] = [0,400]\times [0,200]$ is discretized on an equidistant mesh of $N_x\times N_y = 500 \times 250$ quadratic elements with linear (Q1) ansatz and test functions. The spatial discretization was tested for convergence so that further refinement had no influence on the obtained results. An implicit second order Runge-Kutta scheme \cite{alexander1977diagonally} is used for time-integration. A Newton method solves the resulting nonlinear problem, employing a biconjugate gradient stabilized method (BiCGStab) with a symmetric successive overrelaxation (SSOR) as preconditioner for the linear problems, also see \cite{bastian2010generic,bastian2008genericI,bastian2008genericII} for details of the framework.  
To implement Eq.~(\ref{TFEmodel}) using this ansatz, we split the model into two equations second order in space:
\begin{eqnarray}
 \partial_t h &=& \nabla \cdot \left[\frac{h^3}{3} \nabla [- w - \kappa\Pi(h)] + \boldsymbol{\chi}(h) \right] -\mathbf{U} \cdot \nabla h\label{TFEdns}, \\
 w &=& \Delta h.
\end{eqnarray}
In a weak formulation for general test functions $\phi_h$ and $\phi_w$ the equations read
\begin{eqnarray}
\int_\Omega \partial_t h \phi_h \, \mathrm{d}x\mathrm{d}y =&  \int_\Omega  \frac{h^3}{3} \nabla (w+\kappa\Pi(h)) \cdot \nabla \phi_h \, \mathrm{d}x\mathrm{d}y \nonumber\\
 & - \int_{\partial\Omega}  \phi_h   \frac{h^3}{3}  \nabla \big(w + \Pi(h)\big) \cdot \mathbf{d}\mathbf{s}\nonumber\\
& -\int_\Omega (\boldsymbol{\chi}(h)-\mathbf{U}h) \cdot \nabla \phi_h \, \mathrm{d}x\mathrm{d}y \nonumber\\
& + \int_{\partial\Omega}  \phi_h  \left( \boldsymbol{\chi}(h)-\mathbf{U}h \right) \cdot  \mathbf{d}\mathbf{s},
 \end{eqnarray}
\begin{eqnarray}
0 &=& \int_\Omega w \phi_w + \nabla h \cdot \nabla \phi_w \mathrm{d}x\mathrm{d}y -  \int_{\partial\Omega}  \phi_w  \nabla h \cdot  \mathbf{d}\mathbf{s}.
\end{eqnarray}
The boundary integrals with the exterior unit normal element $\mathbf{ds} = \mathbf{n}\,\mathrm{d}s$ occurring due to the performed partial integrations vanish along the $y$-direction due to the used periodic boundary conditions and are used to implement the boundary conditions (\ref{eq:bc}) in $x$-direction.\par
As initial conditions we use a height profile that linearly decreases from the imposed boundary height $h_0 = 100$ to the height of the wetting layer $h_\mathrm{p} = 1$ with slope $-\alpha$ and then stays constant, i.e., defining a sharp, artificial meniscus. \par
To determine steady profiles in the one-dimensional case for homogeneous substrates we employ path-continuation techniques \cite{Kuznetsov2010,DWCD2014ccp}. These allow one to directly follow solutions in parameter space. They combine prediction steps that advance a known solution in parameter space, e.g., via a tangent predictor, and corrector steps that employ Newton-type procedures to converge to the next solution. In particular, we use the pseudo-arclength continuation package auto07p \cite{DoKK1991ijbc,DoKK1991ijbcb}. To obtain the first meniscus solution for our runs, we start with a homogeneous flat film without inclination or transfer velocity and then subsequently continue in these parameters to the wanted starting values. Then we vary parameters in their ranges of interest, e.g., inclination $\alpha$, transfer velocity $U$ and wettability $\kappa$ to obtain all wanted solutions and corresponding bifurcation diagrams. In the bifurcation diagrams, the contact line position is used as solution measure, which we define as the position of the local maximum of the second derivative of the height profile $h$ with respect to $x$. Also see Fig.~\ref{fig:example_profile} in the Appendix for an illustration.\par
We employ the continuation technique on the one hand to compare the corresponding results with the ones obtained by direct numerical simulations. This provides a good benchmark for both numerical methods as they are rather distinct (finite differences vs.\ finite elements, adaptive mesh vs.\ fixed mesh, stationary problem vs.\ dynamic problem). On the other hand, the approach allows us to predict the dynamics of a meniscus on prestructured substrates using the bifurcation diagram for varying, but homogeneous wettability. In this way we can effectively reduce the complexity resulting from the large variety of possible prestructures and show that our results can be generalized beyond the specific geometries discussed here.

\section{Steady profiles on homogeneous substrates}
\subsection{Bifurcation diagrams for varying velocity and inclination angle}
\label{sec:bifurcation_diagram_v_length}
\begin{figure}[htbp]
\center
    \includegraphics[width=0.5\textwidth]{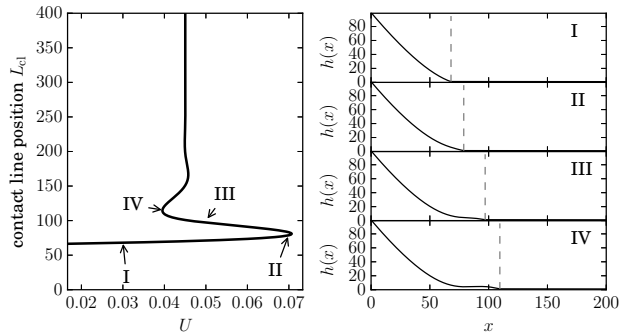}
\caption{Left panel: Bifurcation diagram of a dragged meniscus for varying substrate velocity at fixed wettability $\kappa = 1.0 $ and inclination angle $\alpha = 2.0$. The lower sub-branch corresponds to simple meniscus solutions and ends in a saddle-node bifurcation at an upper threshold velocity. At this bifurcation the sub-branch of simple meniscus profiles connects to one consisting of profiles that show a foot-like protrusion. The length of the protrusion increases when following the upper sub-branch. The right panels give examples of solution profiles at points I-IV marked on the bifurcation curve. The increasing contact line position $L_\mathrm{cl}$ is indicated by vertical dashed lines.}
\label{fig:bifurcation_diagram_v_length}
\end{figure} 

We first discuss the dependence of the contact line position $L_\mathrm{cl}$ on the transfer velocity $U$. As one would expect, $L_\mathrm{cl}$ increases with increasing $U$, because the viscous drag gets stronger and moves liquid further along with the substrate, see the corresponding solution panels I and II in Fig.~\ref{fig:bifurcation_diagram_v_length}. The solutions all correspond to a simple meniscus shape until this sub-branch ends in a saddle-node bifurcation at $U \approx 0.07$ (Fig.~\ref{fig:bifurcation_diagram_v_length}). For larger velocities, instead, a Landau-Levich film of finite thickness much larger than the one of the wetting layer is transferred (not shown here). However, following the second sub-branch that ends in the saddle-node bifurcation for \textsl{decreasing} velocities, we find solution profiles with a foot-like protrusion at the meniscus (see solution panel~III of Fig.~\ref{fig:bifurcation_diagram_v_length}). The foot becomes longer when moving upwards along the snaking bifurcation curve (panel IV, Fig.~\ref{fig:bifurcation_diagram_v_length}) and will eventually span the whole domain, effectively also covering the substrate with a liquid film of finite thickness.\par
This behavior has already been discussed in detail in \cite{galvagno2014continuous,tseluiko2014collapsed}. We briefly recapitulate these findings here, as we employ slightly different boundary conditions and want to emphasize that this has no influence on the general structure of the bifurcation diagram and solution profiles.\par
The main characteristics of the bifurcation curve also persist for different inclination angles. In particular, there exists always a lower sub-branch corresponding to the meniscus solutions ending at a certain limiting velocity, and the snaking sub-branch of foot-like solutions. With other words, one can expect quantitatively different solutions for different inclination angles, but the qualitative overall structure does not change. Hence, without loss of generality, we fix $U = 0.05$ and $\alpha=2.0$ for the following considerations of substrates with different wettability. In Fig.~\ref{fig:bifurcation_diagram_v_alpha_length} in the Appendix, further bifurcation curves are shown over the parameter plane spanned by the velocity $U$ and the inclination $\alpha$.

\subsection{Bifurcation diagram for varying wettability}
\label{sec:bifurcation_diagram_rho_length}
\begin{figure}[htbp]
\center
    \includegraphics[width=0.5\textwidth]{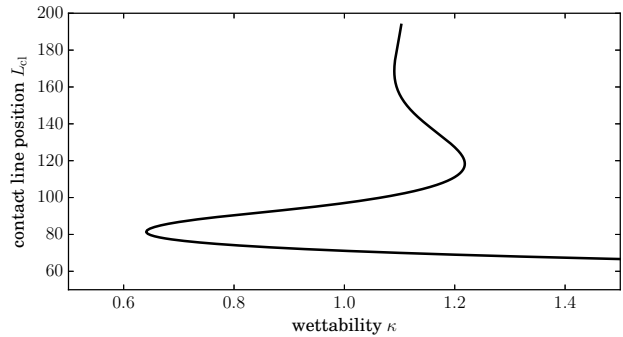}
\caption{Bifurcation diagram of a dragged meniscus showing the contact line position $L_\mathrm{cl}$ depending on the substrate wettability $\kappa$ at fixed inclination angle $\alpha = 2.0$ and velocity $U = 0.05$. The overall shape is similar (but horizontally flipped) to the curve for varying velocity at fixed wettability, cf.~Fig.~\ref{fig:bifurcation_diagram_v_length}. Here, a decreasing value of $\kappa$, i.e., a decreasing contact angle or increasing wettability, takes the role of the increasing velocity in Fig.~\ref{fig:bifurcation_diagram_v_length}.}
\label{fig:bifurcation_diagram_rho_length}
\end{figure}

We now study the bifurcation diagram for varying, homogeneous wettability $\kappa$ keeping all other parameters fixed. As can be seen in Fig.~\ref{fig:bifurcation_diagram_rho_length}, the overall solution structure looks very similar to the case of varying velocity and fixed wettability (cf.~Fig.~\ref{fig:bifurcation_diagram_v_length}) with the major difference that the curve is seemingly horizontally flipped. This indicates that a decreasing $\kappa$, i.e., a decreasing equilibrium contact angle or increasing wettability, has qualitatively the same influence as an increasing transfer velocity. 

As before, there exists a lower sub-branch of meniscus solutions that ends in a saddle-node bifurcation at a limiting lower value of $\kappa \approx 0.64$, below which a Landau-Levich film would be transferred. Increasing $\kappa$ from the bifurcation, again a snaking sub-branch of foot-like solutions is present. We do not show exemplary solution profiles here, as they closely resemble the ones shown in Fig.~\ref{fig:bifurcation_diagram_v_length}. This finding could be expected as a higher wettability means that the contact line region shows less resistance to viscous drag, i.e., can be more easily deformed. However, this has remarkable practical consequences, as it implies that all transitions between different profiles that can be induced by changing the transfer velocity can similarly be induced by changing the substrate wettability. This is particularly interesting, as the wettability of substrates can be tuned by, e.g., chemical processes or by coatings. Both can be applied to produce spatial wettability patterns on the substrates. This in turn implies that when dragging such a heterogeneous substrate from a liquid bath, one can tune the properties of the meniscus in a spatially varying way as well. 

\section{Solutions on heterogeneous substrates}
\label{sec:heterogeneous}
We will now discuss the influence of a spatially varying wettability, i.e., of a chemically prestructured substrate, on the form and dynamics of the liquid meniscus that forms when such a substrate is dragged out from a liquid bath. We discuss two possible cases: (i) a stripe-like prestructure where the stripes are parallel to the transfer velocity, cf. Eq.~(\ref{prestructure_y}), and (ii) stripes that are perpendicular to the transfer velocity, cf. Eq.~(\ref{prestructure_x}). We denote them as (i) ``vertical prestructure'' and (ii) ``horizontal prestructure'', respectively.

\subsection{Vertical prestructure - steady profiles}
\label{sec:vert}
\begin{figure}[htbp]
\center
    \includegraphics[width=0.5\textwidth]{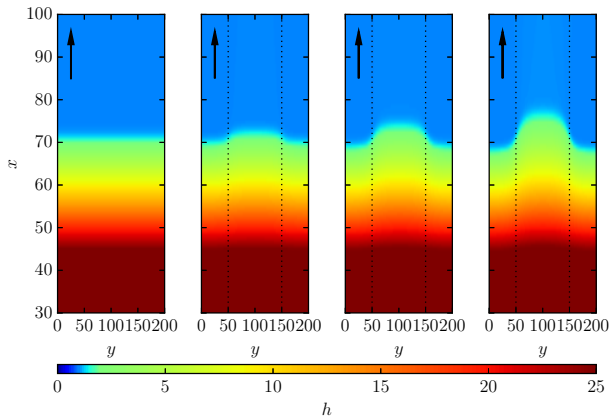}
\caption{Two-dimensional direct numerical simulations of a dragged meniscus on a vertically prestructured substrate with wettabilities (from left to right) $\kappa = 1$, $\kappa = 1 \pm 0.1$, $\kappa = 1 \pm 0.2$, $\kappa = 1 \pm 0.3$ at fixed inclination angle $\alpha = 2.0$ and velocity $U = 0.05$. The dotted lines indicate the boundaries of the prestructure, with the more wettable part at the domain center. For increasing wettability contrast $\rho$, the meniscus gets deformed stronger with a protrusion in the more-wettable region and a receded contact line in the less-wettable region. The black arrows denote the upward dragging direction.}
\label{fig:DNS_vertical}
\end{figure}

We start with the case of a vertical prestructure, i.e., stripes parallel to the transfer velocity described by Eq.~(\ref{prestructure_y}) with a prepattern steepness $s=100$ and wavelength $L_\mathrm{pre} = 200$. Namely, we consider one more-wettable stripe of width $100$ at the center of a domain of size $L_y = 200$. Figure~\ref{fig:DNS_vertical} shows the resulting steady meniscus profiles for different wettability contrasts $\rho~=~0.0,\ 0.1,\ 0.2,\ 0.3$.  

Compared to a homogeneous substrate with $\rho=0$, the meniscus gets deformed when the prestructure is introduced. As one would expect, the liquid is dragged further up the substrate in regions where the substrate is more wettable (at the center  of the domain shown in Fig.~\ref{fig:DNS_vertical}). In the less-wettable region, the position of the contact line is lower than in the unpatterned reference case with $\kappa=1$. The transition between the two regions is smooth, owing to the surface tension of the liquid. When the wettability contrast is increased, the contact line becomes more strongly deformed with the protrusion in the more-wettable region reaching further up. For sufficiently large wettability contrasts (not shown here), the protrusion can finally span the whole $x$-extent of the substrate, effectively corresponding to a stripe-like transfer of a Landau-Levich film onto the more-wettable substrate region.\par
\begin{figure}[htbp]
\center
    \includegraphics[width=0.5\textwidth]{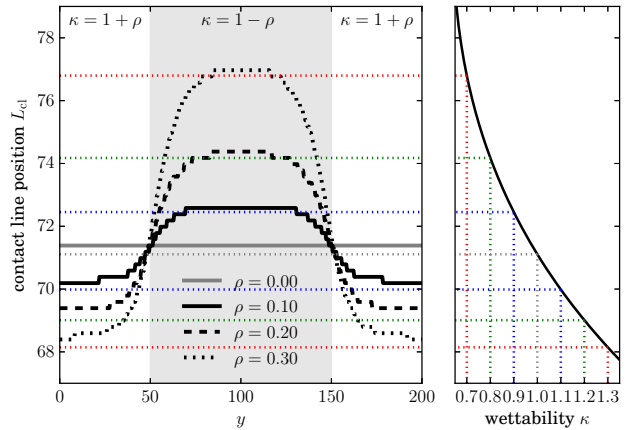}
\caption{Left panel: position of the contact line extracted from the 2d direct numerical simulations shown in Fig.~\ref{fig:DNS_vertical} for various $\rho$ and fixed inclination angle $\alpha = 2.0$ and velocity $U = 0.05$. Right panel: Enlargement of the bifurcation diagram in Fig.~\ref{fig:bifurcation_diagram_rho_length} showing the lower sub-branch corresponding to the 1d simple meniscus solutions. The dotted colored lines are guides to the eye and indicate the contact line positions that one should expect from the 1d calculations on homogeneous substrates with corresponding wettability. The absolute differences between the 1d prediction and the numerical 2d results are all smaller than $\Delta L_\mathrm{cl}=0.4$ with the relative errors being smaller than 0.6\%.}
\label{fig:meniscus_position_space}
\end{figure}

To also quantitatively study the influence of the wettability contrast $\rho$, we again employ the contact line position $L_\mathrm{cl}$ as a measure, which now varies along the $y$-direction,
\begin{equation}
 L_\mathrm{cl}(y) = \arg\max_x \left(\partial_{xx}h(x,y)\right).
\end{equation}
This contact line position is shown in the left panel of Fig.~\ref{fig:meniscus_position_space} for the same parameters as used in Fig.~\ref{fig:DNS_vertical}.

Here, the tendency described above can be seen more clearly. In addition, it is visible that the contact line position reaches plateaus in the middle of the prestructure stripes at $y \approx 100$ and at the domain boundaries. This indicates that the meniscus profile exhibits only small variations in the $y$-direction inside the regions of homogeneous wettability. Accordingly, the transition between the two plateaus is localized at the boundaries of the prestructure. However, for larger prestructure contrast, because of surface tension, this transition regions become wider and the plateaus become smaller. \par
In addition, the right panel of Fig.~\ref{fig:meniscus_position_space} shows a zoom into the bifurcation diagram in Fig.~\ref{fig:bifurcation_diagram_rho_length}, employing the same contact line position axis as the left panel. In this way, the one-dimensional results obtained for homogeneous substrates with the continuation technique can be readily compared to the inhomogeneous two-dimensional case. The dotted colored lines indicate the predictions for the wettabilities $\kappa$ that occur in the direct numerical simulations. For instance, the green dotted lines in the right panel at the wettabilities $\kappa = 0.8$ and $\kappa = 1.2$ correspond to the dashed black profile in the left panel for $\rho = 0.2$. Apparently, the bifurcation diagram for one-dimensional solutions predicts very nicely the position of the plateaus in the contact line position on the prestructured substrate. The contact line position in the direct numerical simulation is always systematically slightly above the values suggested by the bifurcation diagram (deviation $\Delta L_\mathrm{cl}<0.4$). This can also be seen for the case of a homogeneous substrate in the direct numerical simulations (see gray line in left panel of Fig.~\ref{fig:meniscus_position_space}, corresponding to $\kappa=1.0$), possibly owing to the different numerical approaches. However, from this comparison one can see that the results for a particular prestructured substrate can to a large extent be predicted by considering the results for homogeneous substrates of corresponding wettabilities. This is in particular relevant as the latter are inherently one-dimensional and therefore achievable with less effort. In addition, the results suggest that they can be readily extended to other cases, e.g., to prestructures with larger wavelength $L_\mathrm{pre}$. Then, the plateau region will be broader, but their position and the transition regions between them will remain unchanged. Therefore, it is not necessary to calculate various differently shaped vertical stripe prestructures as the results can be easily deduced from the 1d bifurcation diagram.

\subsection{Horizontal prestructure - time-periodic solutions}
\begin{figure}[b]
\center
    \includegraphics[width=0.5\textwidth]{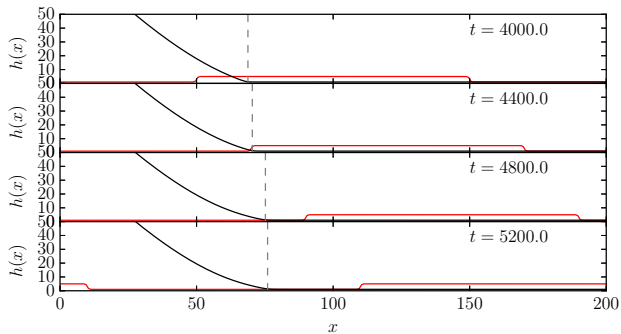}
\caption{Snapshots of a direct numerical simulation of a 1d dragged meniscus on a substrate with horizontal prestructure stripes of wettabilities $\kappa = 1 \pm 0.3$, for fixed inclination angle $\alpha = 2.0$, velocity $U = 0.05$ and prestructure wavelength $L_\mathrm{pre}=200$. The black lines indicate the height profiles, while the red lines indicate the form and position of the wettability profile, where a higher value corresponds to lower wettability. The dotted vertical lines mark the respective contact line positions $L_\mathrm{cl}(t)$.}
\label{fig:DNS_horizontal}
\end{figure} 
Next, we investigate the case of a horizontal prestructure, i.e., stripes perpendicular to the transfer velocity described by Eq.~(\ref{prestructure_x}). In contrast to the previous case where the prestructure was invariant in the direction of the transfer velocity, the prestructure now moves with the transfer velocity $U$. Therefore, now the resulting solutions are no longer steady but periodically vary in time. As now the prestructures are invariant in $y$-direction, we use 1d simulations to determine the expected time-periodic 1d solution profiles. Figure~\ref{fig:DNS_horizontal} shows four snapshots from such a simulation, where also the prestructure is sketched (red lines at the bottom of the panels). As soon as a more wettable stripe passes below the contact line position (marked by vertical dashed lines), the contact line gets dragged further along with the substrate (starting at $t=4400$). After a short time, an almost steady profile is reached (see panels at $t=4800$ and $t=5200$) that persists as long as the more wettable stripe is positioned below the contact line region.

\begin{figure}[htbp]
\center
    \includegraphics[width=0.5\textwidth]{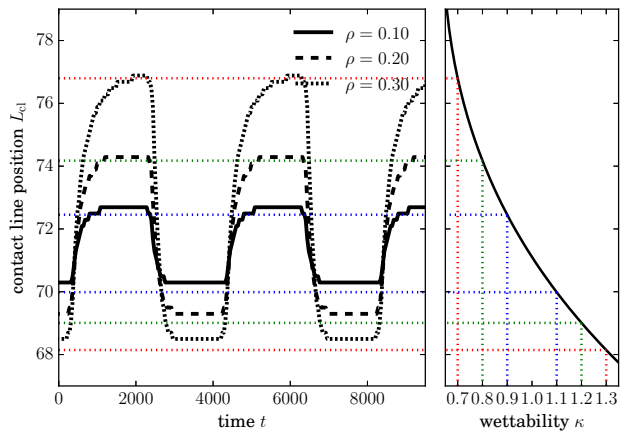}
\caption{Left panel: time-series of the contact line  position $L_\mathrm{cl}(t)$ extracted from direct numerical simulations for a horizontal prestructure with varying contrast $\rho$ and fixed inclination angle $\alpha = 2.0$ and velocity $U = 0.05$. Right panel: Enlargement of the bifurcation diagram in Fig.~\ref{fig:bifurcation_diagram_rho_length} showing the lower branch corresponding to the simple meniscus solutions. The dotted colored lines shall guide the eye to see which contact line positions are to be expected from the one-dimensional, stationary solutions on accordingly wettable homogeneous substrates. The absolute differences between the prediction and the numerical results are all smaller than $\Delta L_\mathrm{cl}=0.4$ with the relative errors being smaller than 0.6\%.}
\label{fig:meniscus_position_time}
\end{figure} 

As before,  we focus our analysis on the contact line  position $L_\mathrm{cl}(t)$, which is now a function of time. The resulting time-series for different prestructure contrasts $\rho$ are shown in the left panel of Fig.~\ref{fig:meniscus_position_time}. On first sight, the results look similar to the case of a vertical prestructure in Fig.~\ref{fig:meniscus_position_space} with the contact line  position smoothly varying between two plateau values, that depend on the prestructure contrast $\rho$. However, we emphasize that Fig.~\ref{fig:meniscus_position_time} shows a time series of positions while Fig.~\ref{fig:meniscus_position_space} presents the static contact line positions in their dependence on the $y$-coordinate. Inspecting Fig.~\ref{fig:meniscus_position_time} more closely, one can identify a left-right asymmetry in the time-series, i.e.., the behavior when entering a more-wettable region is different from the behavior when leaving it (as the symmetry $t \rightarrow -t$ does not hold, while in section~\ref{sec:vert} the symmetry $y\rightarrow -y$ holds). In particular, all transitions between upper and lower plateaus exhibit stronger slopes at the beginning when leaving a plateau value than when approaching the next plateau values. \par
The right panel of Fig.~\ref{fig:meniscus_position_time} again shows the prediction resulting from the bifurcation diagram in Fig.~\ref{fig:bifurcation_diagram_rho_length} for homogeneous substrates. The dotted colored lines indicate the values for the contact line position at the relevant wettabilities. One notes that the plateau values found in the numerical simulations are again well predicted by the bifurcation diagram. Also, the contact line position in the direct numerical simulations is only slightly larger than the one predicted from the bifurcation diagram (deviation $\Delta L_\mathrm{cl}<0.4$).\par
In summary, the results suggest that also in the case of a horizontal prestructure quantitative predictions concerning the time-periodic dynamics of the dragged meniscus can be achieved by consulting the bifurcation diagram for steady solutions on homogeneous substrates.\par

\begin{figure}[htbp]
\center
    \includegraphics[width=0.5\textwidth]{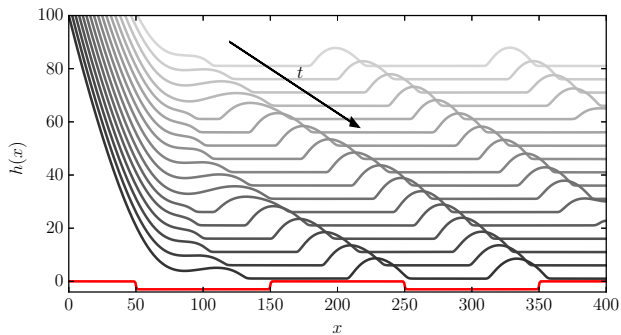}
\caption{Snapshots from a 1d simulation for $\rho = 0.2$, $\alpha = 2.0$ and velocity $U = 0.07$, where the horizontal prestructure  has a contrast that is large enough to partly induce a Landau-Levich film transfer. This leads to the formation of liquid ridges. Note that the employed velocity is higher than in the previous parts. The consecutive snapshots are shifted along the $h$-axis for better visibility with the most advanced state being at the bottom. Note that the distance between the transferred ridges is smaller than the prestructure wavelength $L_\mathrm{pre}=200$, because the ridges are not pinned to the prestructure (indicated as red line).}
\label{fig:stripe_transfer}
\end{figure} 

Yet, this becomes more difficult when considering situations where the wettability difference leads to the transfer of a Landau-Levich film in one of the phases of the dynamics. This is the case, when parts of the substrate have a wettability $\kappa$ lower than the location of the lower fold in the bifurcation diagram (cf.~left fold at $\kappa=0.64$ in Fig.~\ref{fig:bifurcation_diagram_rho_length}). In the case of a horizontal prestructure, this leads to a temporal alteration of the profile between a simple meniscus and a finite height liquid layer, effectively transferring liquid ridges onto the substrate. One example for this case is shown in Fig.~\ref{fig:stripe_transfer}.
Interestingly, the resulting liquid ridge pattern does not necessarily reproduce the spacing of the prestructure perfectly. In the case shown here, the prestructure gradients are too weak to \textsl{pin} the ridges. Instead they are still mobile and can slide downwards with respect to the substrate. Therefore, the distance  between the ridges can be smaller than the wavelength of the prestructure. After the ridges detach from the meniscus, they basically represent 1d drops sliding down an heterogeneous substrate. Hence, results about the (de)pinning of ridges can be readily borrowed from the corresponding literature (e.g., Ref.~\cite{ThKn2006njp}).  

\section{Langmuir-Blodgett transfer onto prestructured substrates}
\label{sec:LB}
We now turn to the Langmuir-Blodgett (LB) transfer of surfactant layers, an experimental system in which the dynamics of a liquid meniscus plays an important role in the pattern formation process \cite{RiSp1992tsf}. For LB transfer, one prepares a monolayer of amphiphilic molecules adsorbed at the free surface of a water reservoir, from which a solid substrate is then withdrawn at a specific transfer velocity. Depending on the transfer conditions, such as the transfer velocity and the surface pressure, either a homogeneous or a patterned deposit of the amphiphilic molecules onto the substrate can be achieved. The resulting patterns consist of two phases of different densities, the so-called liquid-condensed and liquid-expanded phases. The pattern formation in this system has been analyzed to a large extent, both experimentally \cite{RiSp1992tsf,GlCF2000n,LKGF2012s} and theoretically \cite{KGFC2010l,KGFT2012njp,KoTh2014n}, leading to a good understanding of the process. Also the use of prestructured substrates as a way to control the pattern formation has already been analyzed \cite{KGF2011pre,WiGu2014pre,Zhu2016AMI}. However, in these cases prestructures were used or considered which only introduce small wettability changes and thus only have a small influence on the liquid meniscus. In contrast, the prestructures discussed there control the pattern formation process mainly via an influence on the so-called substrate-mediated condensation (SMC) \cite{RiSp1992tsf}, a short-range interaction between the substrate and the amphiphilic monolayer that is responsible for the pattern formation in the monolayer. \par
\begin{figure}[htbp]
\center
    \includegraphics[width=0.5\textwidth]{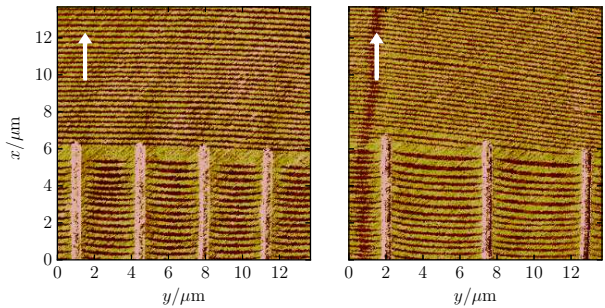}
\caption{AFM phase images of prestructured substrates after LB transfer of DPPC. The horizontal line pattern exhibits an upwards bending towards the gold stripes of a vertically aligned prestructure (left panel: 3\,$\mu$m spacing; right panel: 5\,$\mu$m spacing). The upper halves of both panels show regions of the substrate without prestructure, while the prestructure is visible in the lower half as vertical pinkish lines. The transfer direction is from bottom to top and indicated by the white arrows.}
\label{fig:experimental_bending}
\end{figure}

We now discuss the case where the prestructure also significantly affects the wettability of the substrate and hence the dynamics of the liquid meniscus. In particular, we consider the transfer of L-$\alpha$-dipalmitoyl-phosphatidylcholine (DPPC) onto a silicon (100) substrate with native oxide layer. A 50\,nm thick gold prestructure consisting of 500\,nm wide vertically aligned stripes is produced by electron beam lithography and subsequent vacuum deposition of a 3\,nm chrome adhesion layer and the final gold layer. The substrate is then withdrawn from the bath with a velocity of 32\,mm/min at a liquid temperature of 25\,$^{\circ}$C and a surface pressure of $3.5$\,mN/m. For more experimental details we refer to \cite{Zhu2016AMI} where the same experimental method is used. Figure~\ref{fig:experimental_bending} shows AFM phase images of the resulting patterns for two differently spaced prestructures. On a homogeneous substrate, the pattern consists of horizontal lines perpendicular to the transfer direction (see the upper regions of Fig.~\ref{fig:experimental_bending}).
These horizontal line patterns become interrupted by the vertically oriented prestructure stripes, but persist in the channels between them. However, the lines are no longer straight and horizontal, but exhibit a curvature such that the end of the lines is bent upwards at the sides of the channels between the prestructure stripes. As the formation of the LB patterns takes place at the meniscus, where the distance between the monolayer and the substrate becomes small enough for the SMC to play an important role, this deformation of the lines is a strong indicator that also the contact line position is affected by the prestructure.\par
To describe this experiment, we employ a long-wave model describing the dynamics of the liquid layer forming the meniscus as used before, which is now coupled to the advection-diffusion dynamics of the monolayer floating on top of the liquid layer.  Such a model was previously developed and already proved its capacity to describe the physico-chemical pattern formation processes during LB transfer \cite{KGFC2010l, KGF2011pre}. It consists of two coupled nonlinear partial differential equations for the height $h(x,y,t)$ of the liquid layer and the area density $\Gamma(x,y,t)$ of the surfactant molecules:
\begin{eqnarray}
 \partial_t h(x,y,t) = -\nabla \left[\frac{h^3}{3}\nabla \left( (1-\epsilon^{-2} P_\mathrm{hom}) \Delta h + \kappa\Pi(h) \right)\right. \nonumber \\
\quad \quad \left.+ \frac{h^2}{2} \left( \epsilon^{-2}\Gamma \nabla \Delta \Gamma - \nabla P_\mathrm{hom}\right) -h \mathbf{U}\right] + Q_\mathrm{ev} \label{eq:LB1}\\
 \partial_t \Gamma(x,y,t) = -\nabla \left[\frac{\Gamma h^2}{2}\nabla \left( (1-\epsilon^{-2} P_\mathrm{hom}) \Delta h + \kappa\Pi(h) \right)\right. \nonumber \\
 \quad \quad \left.+ \Gamma h \left( \epsilon^{-2}\Gamma \nabla \Delta \Gamma - \nabla P_\mathrm{hom}\right) -\Gamma \mathbf{U}\right].\label{eq:LB2}
\end{eqnarray}
The first term in both equations corresponds to a generalized pressure gradient, which comprises the typical Laplace pressure term proportional to $\Delta h$ corresponding to surface tension, as well as the Derjaguin (or disjoining) pressure $\Pi(h)$, which reflects the interaction of the free surface of the liquid layer and the substrate. Here, we employ a disjoining pressure of the form $\Pi(h) = \left(\frac{1}{h^3} - \exp(-\chi_0 h)\right)$, which accounts for the polarity of water \cite{Shar1993l}, but the overall dynamics is comparable to the one used in the previous sections. The prestructure on the substrate enters the model through a periodic modulation of $\kappa$, see Eq.~(\ref{prestructure_y}). The second term in both equations represents Marangoni forces, i.e., local variations of the surfactant layer density leading to surface tension gradients affecting the overall dynamics. Here, $P_\mathrm{hom}(\Gamma)$ is the surface pressure of the monolayer in the vicinity of the liquid condensed -- liquid expanded phase transition, where we assume a Cahn-Hilliard-like double well form. This term, in particular, couples the thermodynamics of the surfactant layer to the dynamics of the liquid layer. The third term in the equations is an advection term again describing the drag of the moving plate. Evaporation of the liquid layer is incorporated via the final term $Q_\mathrm{ev}$ in the equation for $h$. We refer to \cite{KGFC2010l, KGF2011pre} for detailed derivations and descriptions of the model. For a discussion in the context of general gradient dynamics models for thin films with insoluble surfactants and dip-coating procedures see Refs.~\cite{ThAP2012pf}  and \cite{WTGK2015mmnp}, respectively.\par
\begin{figure}[htbp]
\center
    \includegraphics[width=0.5\textwidth]{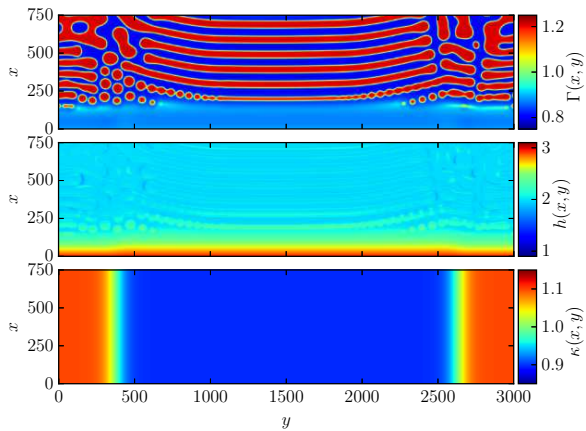}
\caption{Snapshot of a direct numerical simulation of the hydrodynamic model (\ref{eq:LB1}-\ref{eq:LB2}) describing LB transfer of surfactants onto a vertically prestructured substrate. The occurring LB surfactant density $\Gamma(x,y)$ exhibits lines which are bent upwards in the vicinity of the prestructure stripes at the left and right boundary. Further parameters are: $U = 1.3$, $\chi_0 = 1.04$, $\rho = 0.1$.}
\label{fig:LBmodel}
\end{figure}

We now conduct a two-dimensional direct numerical simulation of this model (\ref{eq:LB1})-(\ref{eq:LB2}) using finite differences discretization in space and an adaptive Runge-Kutta 4/5 time-stepping scheme. We employ simple boundary conditions that were also used in Refs.~\cite{KGFC2010l, KGF2011pre}, namely, a fixed low liquid height $h_0 = 3.0$ and fixed monolayer density $\Gamma_0 = 0.9$ and vanishing second derivatives at the downstream boundary. In contrast to the boundary conditions in the previous sections, these conditions are only appropriate in the very vicinity of the meniscus. Corresponding to the experimental setting in Fig.~\ref{fig:experimental_bending}, we consider a simulation domain that spans the distance of two prestructure stripes in horizontal direction, with the prestructure stripes at the right and left hand side and the chanel of bare substrate at the center  (cf.~bottom panel of Fig.~\ref{fig:LBmodel}). 
The surfactant layer density depicted in the top panel of Fig.~\ref{fig:LBmodel} shows that the central horizontal lines exhibit an upward bend towards the edges of the prestructure, i.e., in the direction of the transfer velocity just as in the experiments shown in Fig.~\ref{fig:experimental_bending}. The pattern on the prestructure stripes at the edges of the domain is more irregular. However, this cannot be compared to the experimental results, as there the prestructure stripes are rather rough and do not allow one to identify the phase of the monolayer. As there is a strong coupling between the surfactant layer and the liquid film, the curved line pattern can also be seen in the liquid height profile $h(x,y)$ (see middle panel of Fig.~\ref{fig:LBmodel}). In addition, one can observe that the contact line position is different on and in-between the prestructure. This is similar to the case of the meniscus of simple liquid shown above in Fig.~\ref{fig:DNS_vertical}. Here, however, this has a profound influence on the pattern formation in the monolayer. The patterns are formed where the SMC starts to act, which is at the meniscus, as there the distance between the substrate and the monolayer becomes very small. Hence, the following mechanism can be assumed: a change in wettability of the substrate leads to a change of the contact line  position, which in turn leads to a shift of the positions where the patterns are formed, effectively bending the patterns in the capillary transition regions where the wettability changes.\par
In summary, this example of Langmuir-Blodgett transfer elucidates how the effects of a prestructure on a meniscus of simple liquid discussed in the previous section can have a significant influence in related problems where complex fluids are deposited on a prestructure. Very similar effects may also occur in other comparable experimental set-ups, such as in the dip-coating of solutions or suspensions \cite{wang2015addressable}.

\section{Summary and Discussion}
\label{sec:summary}
We have analyzed dragged meniscus problems for different homogeneous and spatially-varying substrate wettabilities in the framework of a thin-film (or long-wave) model for the cases of a simple liquid and for a liquid covered with an insoluble surfactant as an example of a complex liquid. Using 1d continuation techniques we have studied the possible steady film profiles for different homogeneous substrate wettabilities as well as the corresponding bifurcation diagram. The latter closely resembles a similar diagram for fixed wettability but varying transfer velocity. This is an important finding, as it means that various bifurcation scenarios and possible transitions that can be triggered via changes of the transfer velocity may also be induced by changes of the substrate wettability. Further, we have investigated the dynamics of the meniscus on heterogeneous substrates with stripe-like prestructures of two different orientations. The obtained results have been related to the bifurcation diagram obtained for homogeneous substrates. The established connection provides extensive possibilities for prediction of the dynamical behavior of meniscii dragged along by two-dimensional heterogeneous substrates by using the information provided by the bifurcation diagram for the one-dimensional homogeneous case. This shows,  in particular, that the obtained results are not limited to the specific prestructure geometries discussed here, but may be readily extended to others. This emphasizes the power of our chosen approach of combining continuation techniques to obtain bifurcation diagrams and direct numerical simulations to complete the picture with dynamical information.\par
Finally, we discussed the influence of a prestructured substrate on the liquid meniscus and pattern formation occurring in LB transfer of a surfactant layer from a bath to a moving plate by discussing both, experimental and theoretical results. We have shown that the results for meniscus dynamics discussed before in the case of a simple liquid also remain valid in more complex situations like the LB transfer. In addition, the hydrodynamics of the meniscus can here in turn also affect the pattern formation process of the surfactant layer, resulting in interesting bent line patterns.\par
We note that the investigations conducted here are not exhaustive. There exist more experimentally relevant control parameters that still need to be systematically analyzed, such as the steepness of the prestructure as compared to the capillary length scale, as their ratio should influence, e.g., the transition regions in the vicinity of the prestructure boundaries. The future use of continuation techniques for the full two-dimensional problem could facilitate this.\par
Another aspect that has only been briefly discussed here but is worth to be further examined is the formation of line deposits of a simple liquid (cf.~Fig.~\ref{fig:stripe_transfer}). Remaining questions are, for instance,  how one can control the wavelength of the patterns, how can one achieve pinning of the patterns on the prestructure, which secondary instabilities can occur in the two-dimensional case or whether such deposits of ridges of simple liquid can be achieved even \textsl{without} prestructures. All these questions go beyond the scope of the present contribution and should be answered elsewhere.\par
Finally, we believe that the employed methods and partly also the obtained results should lend themselves to a transfer to related problems and experimental systems. Especially, coating methods such as slot-die and doctor-blade coating techniques \cite{diao2014morphology}, deal with very similar phenomena as in both cases layers of complex liquids of determined height are transferred from a reservoir onto a moving substrate. Therefore, they present ideal candidates for future studies with similar approaches.

\section*{Acknowledgments}
This work was supported by the Deutsche Forschungsgemeinschaft within the Transregional Collaborative Research Center TRR 61 (B2). J. Zhu thanks the China Scholarship Council for financial support. U. Thiele acknowledges funding through GIF project No. I-1361-401.10/2016. We thank Walter Tewes and Sebastian Engelnkemper for fruitful discussions.

\section*{Appendix}
\begin{figure}[htbp]
\center
    \includegraphics[width=0.45\textwidth]{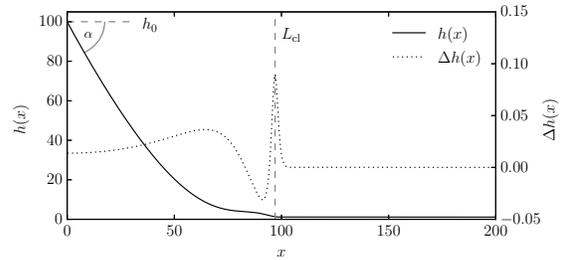}
\caption{Example for a one-dimensional stationary solution of model (\ref{TFEmodel}). The solid black line represents the height profile $h(x)$, while the dotted line indicates the second derivative of the height profile, $\partial_{xx} h(x)$. The maximum of $\partial_{xx} h(x)$ is used to define the contact line position $L_\mathrm{cl}$ (dashed gray line). At the left boundary $x=0$, the profile has the fixed height $h_0$ and slope $-\alpha$.}
\label{fig:example_profile}
\end{figure}

An exemplary one-dimensional steady solution profile is shown in Fig.~\ref{fig:example_profile}. The position of maximal curvature in $x$-direction is used as the definition of the \textsl{contact line position}, i.e., $L_\mathrm{cl} = \arg\max \big(\partial_{xx}h(x)\big)$. \par
\begin{figure}[htbp]
\center
    \includegraphics[width=0.5\textwidth]{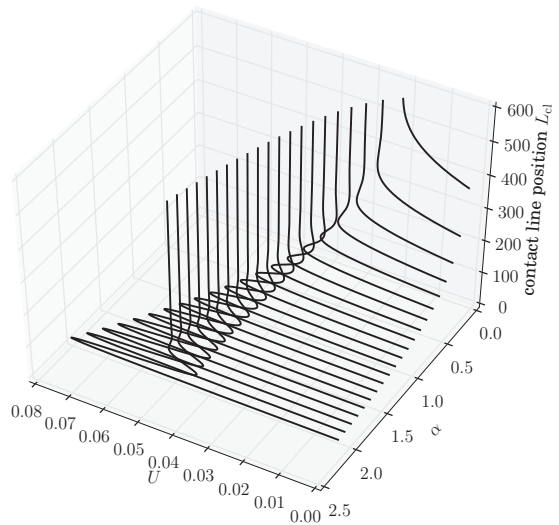}
\caption{Bifurcation curves for the one-dimensional dragged meniscus problem in terms of the contact line position are shown over the parameter plane spanned by the velocity $U$ and the inclination angle $\alpha$. The curves are shown as 'slices' at various fixed $\alpha= [0.2, 0.3, \cdots 2.2]$. For decreasing inclination angles, the contact line position becomes larger and the location of the saddle-node bifurcation, beyond which a Landau-Levich film is transferred, is shifted towards smaller velocities.}
\label{fig:bifurcation_diagram_v_alpha_length}
\end{figure}

Figure~\ref{fig:bifurcation_diagram_v_alpha_length} shows the bifurcation curves as functions of both the velocity $U$ and the inclination angle $\alpha$. Obviously, the main characteristics of the bifurcation curve are similar for different inclination angles. Independent of the inclination, the bifurcation curves always comprise a lower sub-branch corresponding to the meniscus solutions ending at a certain limiting velocity, as well as the snaking sub-branch of foot-like solutions. One can identify a general tendency of decreasing contact line positions and increasing limiting velocity with increasing inclination angles. Note that in the present work we exclude the range of small $\alpha$ below $\alpha\approx0.1$ as there the meniscus solution continuously transforms into a foot-like solution without any saddle-node bifurcation. This transition is analyzed in detail in Refs.~\cite{ZiSE2009epjt,galvagno2014continuous}.
\section*{Literature}
\bibliography{./literature}
\bibliographystyle{unsrt}
\end{document}